# Pigeons trade efficiency for stability in response to level of challenge during confined flight

C. David Williams[1*] and Andrew A. Biewener[1]

1. Harvard University, Concord Field Station, Department of Organismic and Evolutionary Biology, 100 Old Causeway Road, Bedford, MA 01730, USA
* Corresponding author: cdave@uw.edu

# Abstract

Individuals traversing challenging obstacles are faced with a decision: they can adopt traversal strategies that minimally disrupt their normal locomotion patterns or they can adopt strategies that substantially alter their gait, conferring new advantages and disadvantages. We flew pigeons (*Columba livia*) through an array of vertical obstacles in a flight arena, presenting them with this choice. The pigeons selected either a strategy involving only a slight pause in the normal wingbeat cycle, or a wings folded posture granting reduced efficiency but greater stability should a misjudgment lead to collision. The more stable but less efficient flight strategy was not employed to traverse easy obstacles with wide gaps for passage, but came to dominate the postures used as obstacle challenge increased with narrower gaps and there was a greater chance of a collision. These results indicate that birds weigh potential obstacle negotiation strategies and estimate task difficulty during locomotor pattern selection.

# Significance Statement

The real world is a cluttered environment and animals traversing it are faced with innumerable obstacles in their normal locomotion. During normal flight birds have to avoid hitting trees, lamp-posts, and other members of flocks. To investigate flight strategies used in these circumstances we presented pigeons with a simplified challenge, a series of vertical gaps with variable spacings. The pigeons used two discreet postures. One posture granted them greater flight efficiency but was more disrupted when they collided with obstacles, and was used for traversing larger gaps. An alternate flight posture was less efficient but more stable when slight collisions occurred, and was used for traversing smaller gaps. This is the first time we've seen flight strategies tuned during use in cluttered environments.

# Introduction

Locomotion in the real world is uncertain. Error is present in estimations of self-motion and external object location. In environments that present novel or unusually challenging obstacles, moving animals use different strategies to successfully navigate: either maintaining a consistent motor program and relying on passive properties of their neuromechanical systems to carry them through safely, or altering their motor program to detect and avoid or navigate the challenge. Cockroaches traversing substantial obstacles maintain a constant motor pattern (1); whereas, in contrast, blind Mexican cavefish halt their swimming and glide to avoid collision with solid obstacles (2). In avian flight this choice has been subject to little study, despite the confined and cluttered environments in which nesting, social, foraging, and hunting flight behaviors occur. How are birds able to roost in trees and bushes, perch in the gaps of chain-link fences, and fly at speed through forests during hunting without collision? Is this managed through careful path planning while maintaining a constant gait or through altering wingbeat patterns on encountering each new obstacle?

Bird navigation has been studied extensively on the scale of migration with visual, olfactory, magnetic, and auditory map senses reported (3-5). However, local navigation through cluttered environments has received little attention (6), with most local bird flight studies addressing questions of take-off or landing, visual perception, flight kinematics, or of energetics (7-10).

Here we challenge pigeons (*Columba livia, n=4*) to traverse a linear array of vertical poles at a variety of spacings, with gaps ranging from as narrow as 0.2 wingspans to as wide as 0.4 wingspans. While we expected to see a clear traversal strategy emerge as in the terrestrial and aquatic systems noted above, we instead saw the use of two distinct postures, each used by all individuals (Fig. 1). One posture, requiring less disruption of the wingbeat flight pattern, was used predominately to pass though wider gaps, while use shifted to the more disruptive posture as gap size decreased (Fig. 2). We examine the advantages and disadvantages of each traversal strategy that cause them to be selected or discarded when facing a given locomotor challenge.

# Results

**Two distinct postures used - wing pausing and wing folding**

While a variety of navigation methods were expected, two distinct and stereotyped postures (with some overlap in posture metrics, depicted in Fig. 1a-c and S1) were used by pigeons when traversing vertically spaced obstacles. The wings paused posture was achieved by holding the wings static at the top of upstroke, resuming the wingbeat cycle once past the obstacles. The wings flexed or folded posture was used in traversal by interrupting the wingstroke at any point, and retracting the wings closely to the body. Wings held in the paused posture are elevated higher and at a steeper pitch than those in the folded posture (Fig. 1b,c). The wings folded posture is similar to that adopted by many passerines during intermittent bounding flight, although pigeons do not exhibit such flight behavior when flying in open spaces (11). Distinct from wing position during bounding flight, however, in the wing folded posture the pigeon's shoulder remained elevated, while the elbow was flexed and the handwing retracted (adducted) against the body to narrow wing profile breadth.

**A shift from wing pausing to folding as gaps shrink**

The distribution of traversal methods used by the individual pigeons changed with the gap size between the obstacles they were navigating. In general, ignoring the effect of gap size, the paused posture was employed more often than the folded posture. When gaps were wide (approximately 0.4 wingspans) the paused posture was used in almost all trials, but when gaps became narrower and more challenging the folded posture became the more common strategy (Fig. 2). This shift to the folded posture in narrower spacings continued gradually until, at the most challenging spacing which the pigeons would traverse, the folded posture was used in over 60% of trials.

**Paused posture disrupts flight less - shorter duration and less height lost**

That the paused posture is preferentially used when traversing wider, less challenging gaps suggests that it confers benefits over the folded posture; several flight metrics support this (Fig. 3, Table 1). To assess this, we constructed models which looked at the effects on flight metrics of the traversal method used and gap size, in addition to which pigeon was completing each trial. Trial flight metrics included: 1) the time it takes to complete a maneuver, 2) the height lost during and immediately after the maneuver, 3) the speed at which the obstacles were approached, 4) the velocity of the wingtip on the first downstroke after passing through the gap, and 5) the projected position of the wings within a stroke on pass though, had no interruption occurred.

Traversal method had a significant ($p_{effect}$=0.039) effect on the time for which a posture is held, with the wings paused method taking less time than the wings folded method (Fig. 3a, Table 1). Gap size had no significant effect on maneuver duration ($p_{effect}$ = 0.197).

A longer interruption of the wingbeat cycle may be expected to result in more height lost during the traversal. This expectation was borne out; trials in which the folded posture was used lost a mean of 17 cm more height than in trials in which the paused posture was used (p = 0.019, Fig. 3b, with some disagreement between analysis methods discussed in Table 1).

Both maneuver duration and height lost could be expected to vary with the speed of the individual passing through the gap: faster speeds making for faster transits requiring shorter interruption of the wingbeat cycle and consequentially less interruption of lift generation. This would suggest that the advantages of the wings paused posture seen in maneuver duration and height lost were actually due to differences in pre-obstacle flight speed, rather than due to a direct effect of the traversal method on the listed metrics. In practice, pre-obstacle flight speed was not significantly affected by traversal method ($p_{effect}$ = 0.440, Fig. 3c, Table 1), failing to contradict the demonstrated dependence of the prior flight metrics on traversal method and instead supporting the idea that the wings paused posture is itself less disruptive of flight.

Interestingly, pre-obstacle flight speed depended upon gap size ($p_{effect}$ < 0.001, Fig. 3c, Table 1). This suggests an element of caution or apprehension is present in flight control, with more challenging trials being approached and traversed more slowly than those with a wider margin for error before a collision occurs.

In addition to the shorter interruption of the wingbeat cycle, the wings paused posture may allow the initial downstroke after passing though the obstacle to be stronger, as the wings are pre-positioned for downstroke initiation. This effect was not seen, however, with only gap size having a significant effect on the downstroke velocity following the passage through the obstacle array (Fig. 3d, Table 1). This, too, fails to contradict the idea that the wings paused posture itself is less disruptive of flight and lift generation.

**No significant effect of projected wingstroke position**

A further alternate hypothesis to explain the posture choice would be that selection is governed by an individual's position in the wingbeat cycle during obstacle approach. Plausibly, the paused posture may be used in cases where the wingbeat cycle would have been at or near the top of upstroke during obstacle traversal and the folded posture preferentially selected where the wings would be near the bottom of downstroke or transitioning into upstroke. However, the predicted position within the wingbeat cycle (as a fraction of the complete upstroke-peak to upstroke-peak wingbeat remaining) is not significantly different between the traversal methods ($p_{effect}$ = 0.338, Fig. 3e, Table 1). This supports that decision factors other than location within the wingstroke are responsible for the posture used.

**Folded posture is more stable in the face of collisions**

At the narrow spacing at which the wings paused posture was used, occasional contact between the wings and the poles was observed as a result of turn-induced roll from the maneuvering required to center the bird in the gap (11% of trials, predominantly at the smallest two gap sizes, insufficient to block traversal in all but 1.8% of trials). This contact suggests that the wings folded posture used to navigate narrower spacings provides some collision-related benefit not captured by the flight metrics reported above: either by reducing collisions with obstacles in challenging cases, or by mitigating the consequences of collisions when they do occur. Measurements of the distance between wingtip markers indicate that the average stereotyped wings paused posture actually has a narrower profile than that of the folded posture: 16 cm in the paused case as compared to 18 cm in the folded case. Although stereotyped profiles do not account for the narrowest achievable width under either posture, this does not support use of the wings folded posture as a means to reduce the frequency of collisions through a reduced forward profile.

However, modeled collision responses suggest that the folded posture is less susceptible to perturbation than is the paused posture. To analyze this, responses to collisions were quantified with rigid body models of the two postures (Fig. 4a&b) using digitized marker locations and previously described mass distributions (9). All observed collisions occurred along the leading edge of the distal section of the handwing, no head, body, or proximal wing collisions were observed. In colliding areas of the handwing, a perturbing force induced a rotational acceleration approximately 20% higher in the paused posture than in the folded posture (Fig. 4c). The reduced response to a perturbation by the folded wing posture suggests its use when narrower gaps are present may serve to reduce the consequences of unintended collisions with obstacles and increase flight stability.

# Discussion

When faced with an obstacle an animal may alter their gait as they attempt to traverse it. Here, pigeons traversing challenging vertical obstacles choose between a traversal strategy of greater efficiency and a strategy that grants greater stability when a misjudgment leads to collision. Sensibly, the efficient flight strategy is chosen where gaps are wider and there is less chance of a collision occurring, while the more stable strategy is used only as the probability of a misjudged traversal increases.

That the folded posture is the less efficient of the two employed is suggested both directly by the increased maneuver duration and height lost in paused traversals, and indirectly by the aerodynamic measurements made of pigeons and other birds in intermittent flight experiments in wind tunnels where flapping was momentarily ceased (11-12). Pigeons do not engage in intermittent bounding flight with a wings folded phase in free flight, either in the wild or in wind tunnels (11). This is likely due to the decline in bounding flight's efficiency advantage as body size increases: by the time a bird reaches the size of a pigeon there is little or no energy savings from bounding flight (13). This adverse scaling of bounding energetics is attributed to the mass-specific power needed in bounding fight; larger fliers lack sufficient reserve capacity to generate the larger lifts needed to effect and realize an energetic savings from periods of ballistic flight. That pigeons lack the lift reserves needed to benefit from bounding flight is further supported by the lack of difference in wingtip downstroke speed between traversal methods. As a proxy for pectoralis force, this suggests that there is no reserve power to generate more lift and halt the height loss associated with the longer maneuver that results from the interruption of lift generation seen in the wings folded posture (14). Thus the folded posture is the energetically unfavorable approach to obstacle traversals.

However, while the folded posture lacks the efficiency of the paused posture, the primary advantage of the folded posture appears to lie in its increased resistance to perturbation when obstacles being traversed collide with distal wing sections. This increased resistance to perturbation or stability, as shown in our rigid body model, is likely due to the folded posture's combination of 1) a decreased lever arm upon which a collision force can act, and 2) a shallower angle of the collision relative to the surface of the wing allowing more of the force to be shed. These stability advantages explain the selection of a flight maneuver whose opportunity cost is decreased efficiency, and suggests an element of risk assessment where the known decrease in efficiency is balanced against the estimated likelihood of collision.

The choice between these two postures seems mediated by an element of caution or uncertainty. We interpret the approach speed as a proxy indicator of caution or uncertainty, as decreased flight speed permits more time to plan passage through the obstacles but requires increasing amounts of energy as speed decreases. This energy increase is particularly sharp in the sub-5 m/s speed regime used across all obstacle approaches (10). It is likely that the significant decrease in approach speed as trial challenge increases indicates an increased level of caution on the part of the birds, caution that informs the choice of traversal posture.

In these obstacle traversal flights pigeons demonstrated that, when faced by a challenge, an individual selects more minimal or more substantial interruption of their locomotion pattern and that the level of challenge may mediate the decision.

Uncertainty, likelihood of collision, and negative consequences of failed traversals are all factors that drive locomotion strategies towards more cautious choices while the decreased efficiency or speed of cautious choices push towards unaltered locomotion.

# Methods

### Pigeon care

Four wild-caught pigeons were housed in a communal coop at the Concord Field Station (Bedford, MA) in accordance with protocols approved by Harvard University's Institutional Animal Care and Use Committee. When used in flight trials, birds were transported to experimental facilities in individual carriers and provided food and water *ad libitum*.

### Flight corridor and obstacles

A 24 m long by 2.3 m wide by 2.3 m tall flight corridor was used for these trials (Fig. S2). The flight corridor's walls were hung plastic sheeting with which the birds avoided contact or collisions. The ceiling was constructed of stretched 2 cm netting to permit illumination and high-speed filming of flight from above without allowing birds to pass above obstacles in the flight corridor.

An obstacle array was placed in the center of the flight corridor, consisting of a line of evenly-spaced vertical PVC poles. These obstacle poles reached from the floor to the ceiling of the corridor and were 4.2 cm in diameter. The poles were fixed to an extruded aluminum rail (80/20 Inc., Columbia City, IN) at floor level, allowing the space between adjacent poles to be increased or decreased between trials. Thus the birds were obliged to traverse obstacle arrays containing gaps between 13 and 26 cm, approximately 0.2 to 0.4 wingspans for the mean 64 cm wingspan of birds used in this study. Random permutations of gap sizes were used from trial to trial, minimizing anticipation and learning.

Birds were trained to fly on cue between 1.2 m tall perches placed at either end of the corridor. Flight obstacles were located a minimum of 8.5 m from either perch, causing the birds to consistently approach the obstacles in even, level flight. No significant difference in flight path height or angle was observed across trials for individual pigeons. In all gap sizes used, pigeons voluntarily traversed the obstacle array in more than 95% of the attempted trials. Trials in which the bird refused to traverse the obstacle array were discarded. Use of gap sizes below 0.2 wingspans resulted in sharply increased refusal rates.

### Filming and markers

Throughout all trials (n=78), birds were recorded with three FastCam SA3 and two FastCam 1024PCI cameras (Photron USA Inc., San Diego, CA). Synchronized video was recorded at 500 Hz with a shutter speed of 1/1000 s. Videos were saved and evaluated as uncompressed AVIs. Excerpts from these videos are shown in Fig. S1.

Custom active markers (eight sub-0.1g surface-mount infrared LEDs) were mounted to key anatomical locations to speed video processing (Fig. S3a). These active markers appeared to the infrared sensitive cameras as bright point sources, but did not attract the attention of the pigeons to which they were mounted. A backpack-mounted lithium battery system powered these markers. The combined backpack and marker system never exceeded 9% of an individual's mass. Active markers were placed at the center of the head, 2 cm from the left and right wingtips, at the left and right wrists, and in a plane on the bird's back surrounding the powering battery. Marker locations were reconstructed using DLTdv5 (15) and analyzed using custom MATLAB (Mathworks Inc., Natick, MA) and Python scripts. Raw point clouds from this digitization are shown segregated by posture in Figure S3b. When tracking markers over multiple frames, marker location was filtered by a fourth-order Butterworth filter with a low-pass cutoff frequency of 100 Hz, more than twice wingbeat frequency.

**Trial evaluation and modeling of effects**

On viewing flights at reduced playback speed, two distinct postures were used to traverse the obstacle array: 1) a paused posture where the wings were held at the top of the wingstroke, and 2) a folded posture where the hand-wing was hinged back, adducting at the wrist to reduce the wingspan.

Each trial's posture was first visually coded as paused or folded after examining multiple views (by one observer: CDW). Following this, sufficient points were visible for digitization during the execution of a posture to quantitatively classify 88% of trials by two posture metrics: 1) wing pitch relative to the torso's backpack markers and 2) elevation of the wrist markers above the torso's backpack markers (Fig. 1c). These two metrics were segregated into groups using a Gaussian or normal mixtures model in statistical analysis software JMP (SAS Institute Inc., Middleton, MA). Treating the trials' locations in pitch-elevation space as a mixture of normal distributions permits the small intersection of the folded and paused populations seen in Fig. 1c. We treat a >50% affinity for a given group as assignment into that group. This assignment method, when applied to the 88% of trials that we quantitatively scored, agreed with visual coding in all but 12% of cases. Further analysis use the normal mixture posture classification for the 59 trials where it is available and the (validated as largely agreeing) visual classification for the remaining trials in which markers were partially occluded.

Referenced metrics of flights are calculated from tracked marker data in the following fashion. Flight speed was taken from differentiation over time of the location of the head mounted active markers. Wingstroke interruption was determined by marking the times at which the wingtip markers were observed to first cease and then resume the stereotyped pattern of the wingbeat cycle. The point at which the birds passed the obstacles was taken to be when the head marker location passed through the plane defined by the obstacle poles. Downstroke wing speed was taken from the maximum velocity of the wingtip marker on the first downstroke after resumption of flapping, although this metric was only taken from the subset of trials (80% of all) in which the view of the wingtip markers was not occluded by obstacles. The wingbeat cycle was determined by selecting the points at which the wings were at the peak of upstroke and the bottom of downstroke. The cycle duration was found by taking the mean of the resulting up and downstroke durations. The position in this cycle at which a pigeon would be during obstacle traversal, had it not interrupted its wingbeat cycle, was found by projecting the mean cycle forward from the last upstroke or downstroke completion prior to obstacle traversal.

For flight metrics, effects were screened for significance using standard least-squares linear models in JMP. Models were treated as three-way analyses of variance with interaction subject to the fixed factors: individual pigeon, gap size, maneuver method, and the interaction between gap size and maneuver method. Model effects were deemed to have significantly affected a flight metric based upon F-ratio and corresponding p values (< 0.05 taken to be a statistically significant effect size). Model results are summarized in Table 1 and Figure 3.

Stability estimations shown in Figure 4 were calculated as the instantaneous angular acceleration resulting from a force applied to a point along a rigid body model of the pigeon.

$$\frac{d}{dt}\omega(t) = I(t)^{-1}(r_i(t) \times F_i(t)) \tag{1}$$

The mass locations in the rigid body model were calculated from marker positions (Fig. S3a) and the mass magnitudes were taken as fractions of the individual's body mass measured directly preceding flight trials, both as in Ros *et al.*, 2011.

## Acknowledgements
Funding was provided by NSF-1202886 to CDW and ONR-BAA 10-002 MURI to AAB.

The authors would like to thank Ivo Ros for helpful discussions and assistance with animal handling.

## Author contributions
CDW and AAB conceived the experiments, CDW and AAB performed the experiments, CDW analyzed resulting data, CDW and AAB wrote the paper.

## Competing Financial Interests
The authors declare they have no competing financial interests.

# References

1. Sponberg S, Full RJ (2008) Neuromechanical response of musculo-skeletal structures in cockroaches during rapid running on rough terrain. *J Exp Biol* 211: 433-446.
2. Teyke T (1985) Collision with and avoidance of obstacles by blind cave fish *Anoptichthys jordani* (Characidae). *J Comp Physiol A* 157: 837-843.
3. Hagstrum JT (2013) Atmospheric propagation modeling indicates homing pigeons use loft-specific infrasonic 'map' cues. *J Exp Biol* 216: 687-699.
4. Mora CV, Ross JD, Gorsevski PV, Chowdhury B, Bingmanet VP (2012) Evidence for discrete landmark use by pigeons during homing. *J Exp Biol* 215: 3379-3387.
5. Wiltschko R, Wiltschko W (2009) Avian Navigation. *Auk* 126: 717-743.
6. Lin HT, Ros IG, Biewener AA (2012) A bird's eye view of path planning: Is there a simple rule for flying in a cluttered environment? *SICB Annual Meeting* 41.1.
7. Robertson AMB, Biewener AA (2012) Muscle function during takeoff and landing flight in the pigeon (*Columba livia*). *J Exp Biol* 215: 4104-4114.
8. Bhagavatula PS, Claudianos C, Ibbotson, M.R. & Srinivasan, M.V. (2011) Optic flow cues guide flight in birds. *Curr Biol* 21: 1794-1799.
9. Ros IG, Bassman LC, Badger MA, Pierson AN, Biewener AA (2011) Pigeons steer like helicopters and generate down- and upstroke lift during low speed turns. *Proc Natl Acad Sci USA* 108: 19990-19995.
10. Tobalske BW (2007) Biomechanics of bird flight. *J Exp Biol* 210: 3135-3146.
11. Tobalske BW, Dial KP (1996) Flight kinematics of black-billed magpies and pigeons over a wide range of speeds. *J Exp Biol* 199: 263-280.
12. Tobalske BW, Peacock WL, Dial KP (1999) Kinematics of flap-bounding flight in the zebra finch over a wide range of speeds. *J Exp Biol* 202: 1725-1739.
13. Rayner J (1985) Bounding and undulating flight in birds. *J Theo Biol* 117: 47-77.
14. Warrick DR, Dial KP (1998) Kinematic, aerodynamic and anatomical mechanisms in the slow, maneuvering flight of pigeons. *J Exp Biol* 201: 655-72.
15. Hedrick TL (2008) Software techniques for two- and three-dimensional kinematic measurements of biological and biomimetic systems. *Bioinspir Biomim* 3: 034001.

# Figures

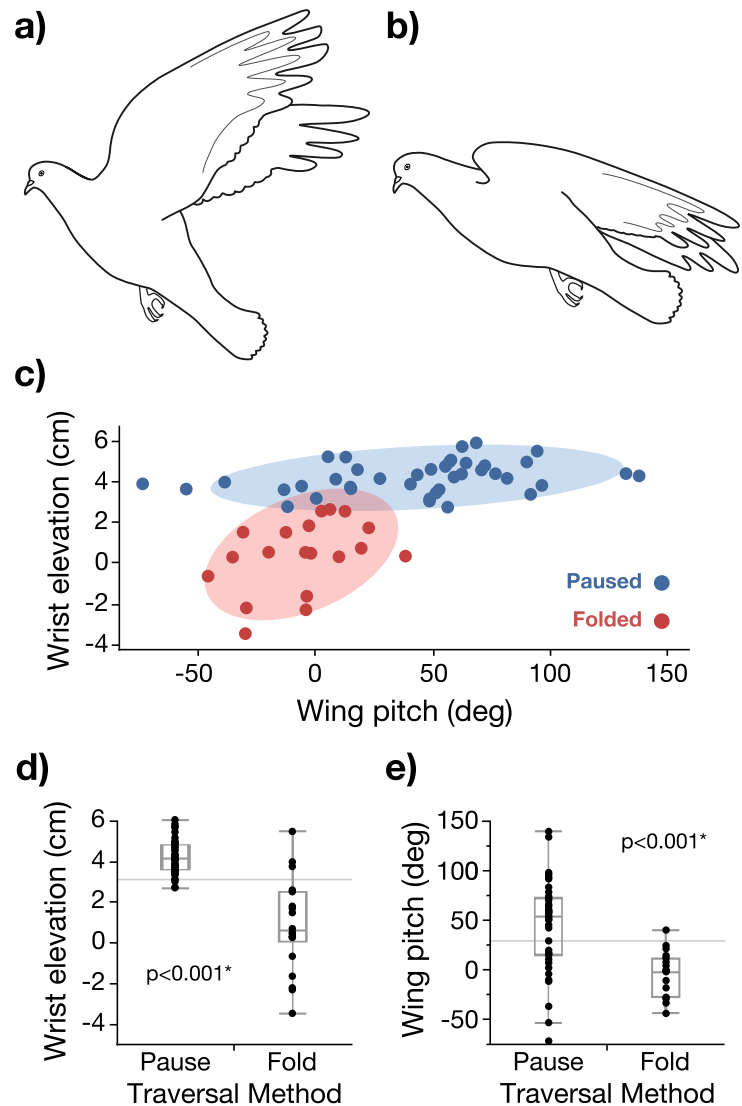

**Figure 1**
**Two traversal postures: wings paused in upstroke and wings folded** - Two stereotyped groups of postures were adopted in traversing the gaps between obstacles: **a)** the "wings paused" case in which the wings were held at the top of upstroke, with downstroke resuming on the far side of the obstacle array and **b)** the "wings folded" case in which the wingstroke was interrupted and the wings tucked along the body and unfolded after passing between obstacles. These postures were quantitatively classified into paused and folded clusters **c)** by applying a normal mixtures model to wrist elevation and wing pitch. Each posture class exhibits a stereotyped or typical posture near the centroid of their respective clusters, within a distribution of similar poses. In **d)** wrist elevation, the height difference between the mean of the torso-mounted markers shown in Fig. S3a and the wrist marker, is directly compared between postures. Similarly in **e)**, wing pitch, the pitch angle between wrist-wingtip line and upper-lower torso markers line, is directly compared between postures. In both cases one-way ANOVA tests show significant differences between each postures' mean pitch and elevation. Additionally, standard least-squares linear models of pitch and elevation as they depend on individual pigeon, traversal method (pause or fold), gap size, and the interaction between traversal method and gap size show traversal method as having a significant effect ($p_{effect} < 0.001$ in both cases).

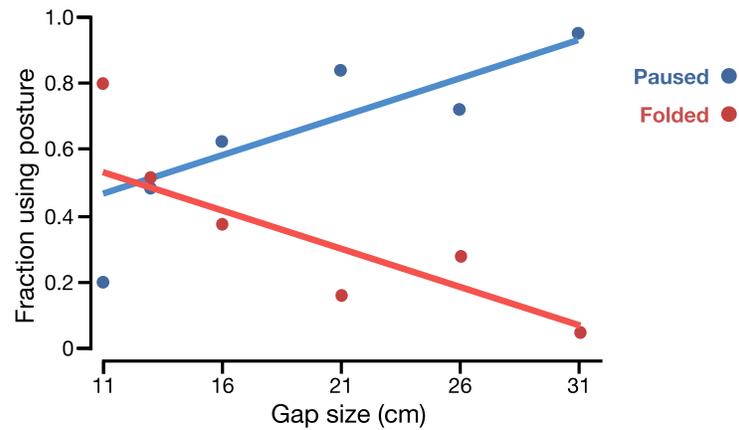

## Figure 2
**Fraction of trials employing each posture as gap size changes** - As the openings narrow, the strategy employed shifts from predominantly pausing the wings to folding the wings in a majority of trials. A least squares linear fit to the fraction folded or fraction paused, means across individuals, shows a significant dependence on gap size (p=0.035 in both cases, as each is the inverse of the other).

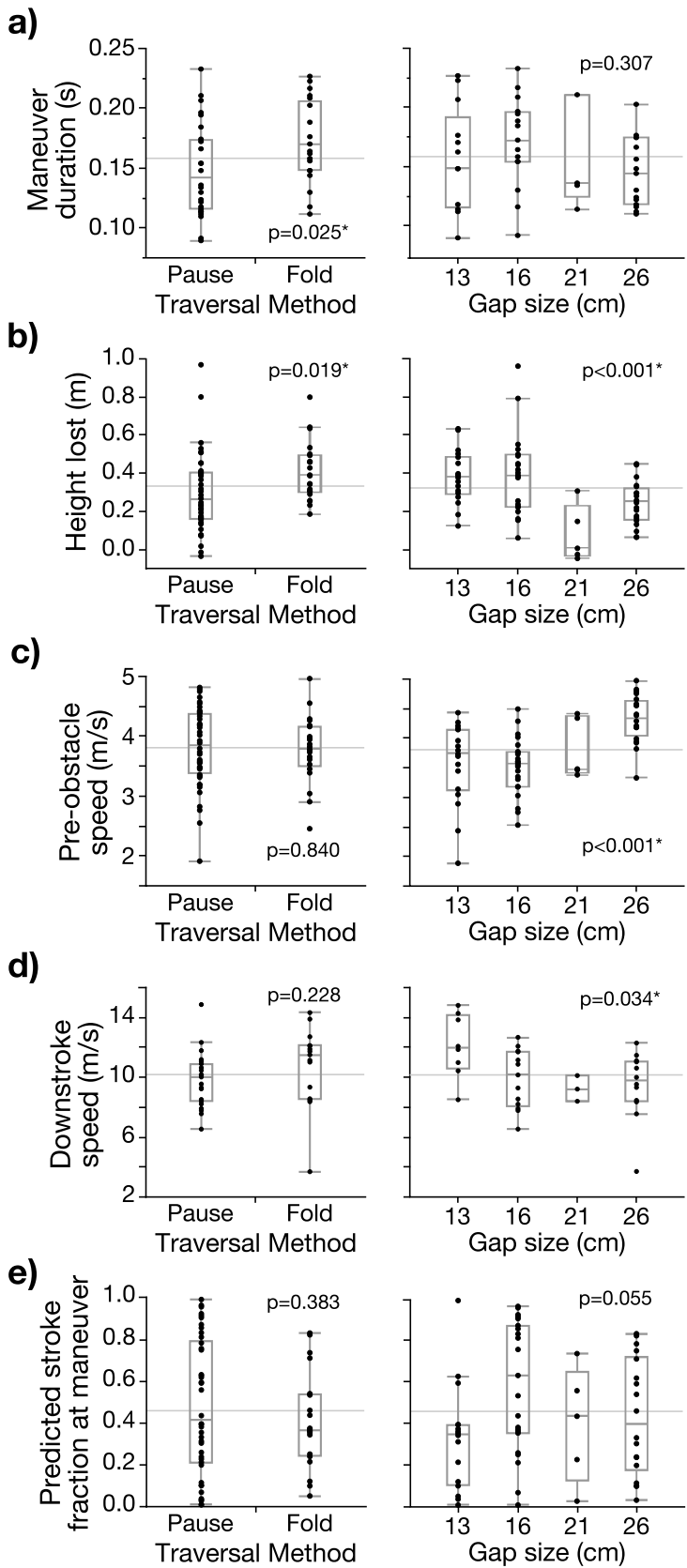

**Figure 3**
**Effects of traversal method and gap size on flight metrics** - Flight metrics are affected by individual pigeon used, traversal method, and gap size. The interactions between traversal method, gap size and flight metrics are shown here as quantile plots (with outlier whiskers) while their significance is judged by ANOVA. Traversing the obstacles with the paused posture **a)** takes less time than doing so with the folded posture and **b)** loses less height than doing so with the folded posture. These advantages of the paused posture are not a result of either **c)** faster traversal of the obstacle array when the paused posture is used nor of **d)** faster downstroke speed when the paused posture is used; neither of these flight metrics are significantly affected by traversal method. Likewise, maneuver duration and height lost are not correlated with gap size. This rules out preference for a given traversal method at a given gap size giving a false correlation between the traversal method and the respective flight metric. There is **e)** no significant interaction between where in the wingbeat cycle a bird would be at pass-though and what traversal method it uses during pass through. Here downstroke starts at 1 and upstroke is completed at 0, giving position within the wingbeat as the fraction of a complete stroke remaining. This, along with the lack of correlation between gap size and stroke fraction, suggests that stroke fraction contributes little to the choice of traversal method.

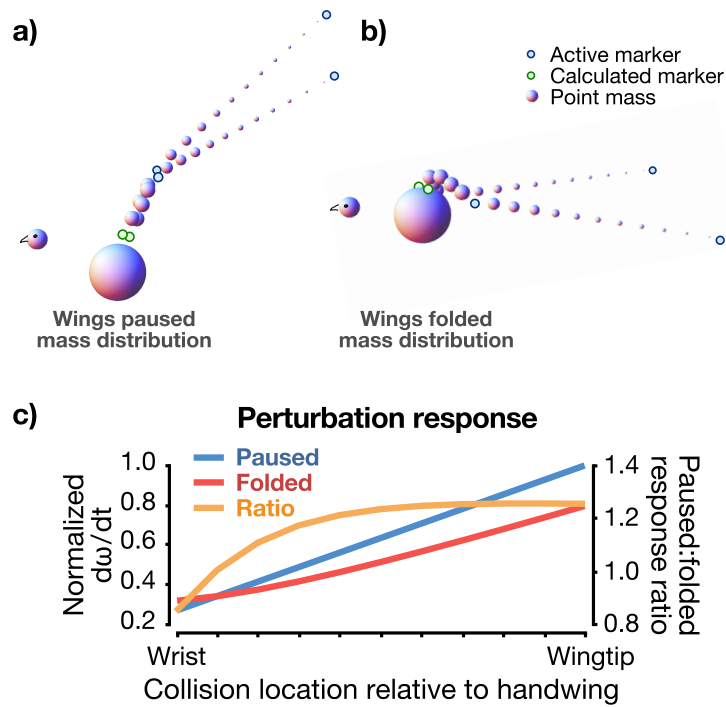

### Figure 4
**Stability comparison between postures** - Positions of the body, the head, and along the wings (as in Fig S3a) determine the locations of point masses in a mass distribution model (9). Spheres whose volume is proportional to their relative mass depict both posture models, with beak and eye markings for orientation. In the paused posture **a)** the mass of the wings is extended above and outward from the body. In the wings folded posture **b)** the wings are tucked lower and closer to the body. The posture models are used to calculate **c)** the response to a perturbing torque generated by a simulated forward flight collision between the environment and the handwing (where all observed obstacle collisions were observed to impact). This response is given as the angular acceleration (dω/dt) resulting from a perturbing torque, normalized to the maximal acceleration, providing a measurement of destabilizing rotation to be countered following wing strikes. The folded posture suffers less rotational perturbation across 90% of the wingspan, reflecting a more stable posture.

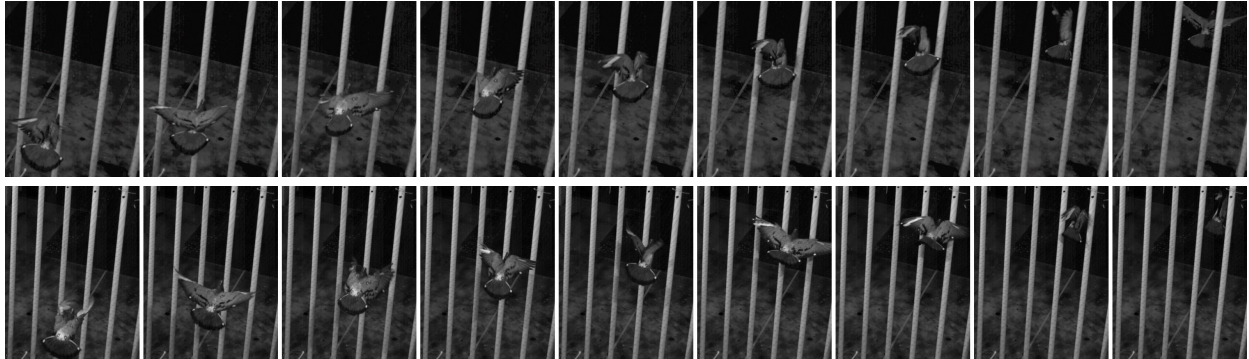

**Supplemental Figure 1**
**Two traversal postures** - The wings paused and wings folded postures are shown in filmstrip excerpts from these two traversals with 16 ms passing between each frame. In the upper trial, the wings were paused at the top of the upstroke, remaining fixed above the pigeon during traversal. In the lower trial, the wings were folded back and held before passage through the obstacle array.

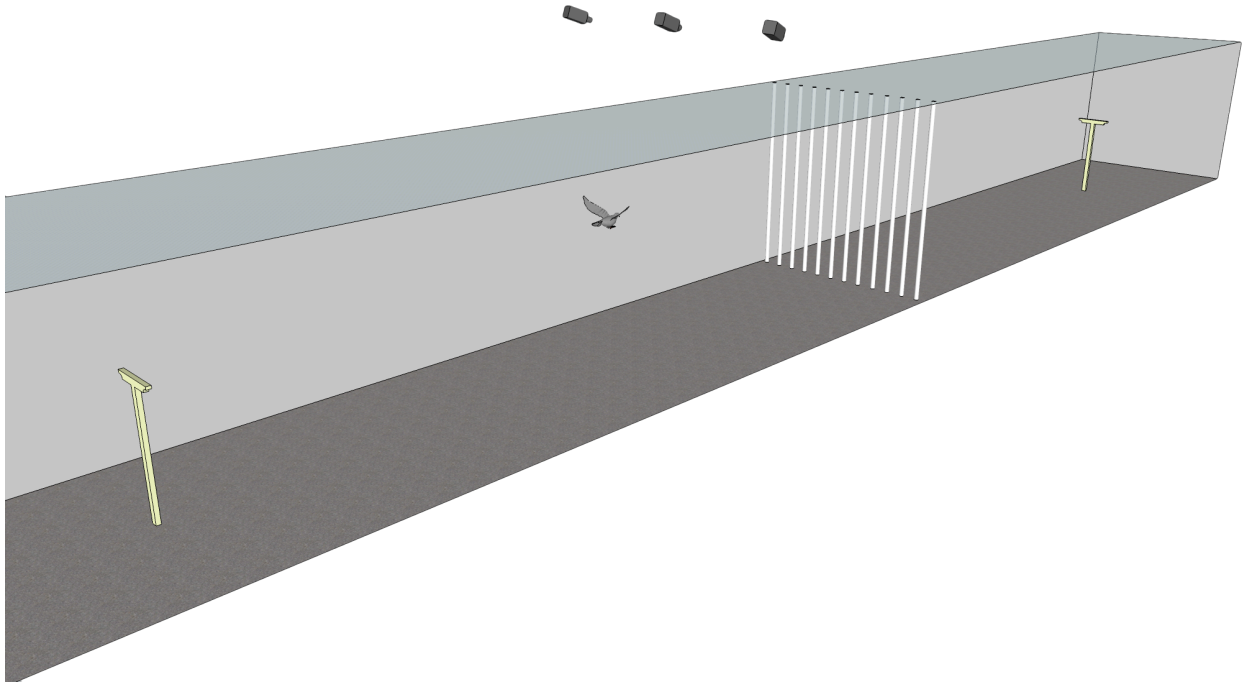

## Supplemental Figure 2
**Flight corridor and obstacle array** - A 24 m long by 2.3 m wide by 2.4 m tall flight corridor was used for these trials, with a linear array of obstacles located 8.5 m from perches at either end of the corridor. High-speed cameras were mounted 1.4 m above the obstacle array and netting, which formed the corridor ceiling. Pigeons were trained to fly from perch to perch in the absence of obstacles and continued to do so once obstacles were put in place.

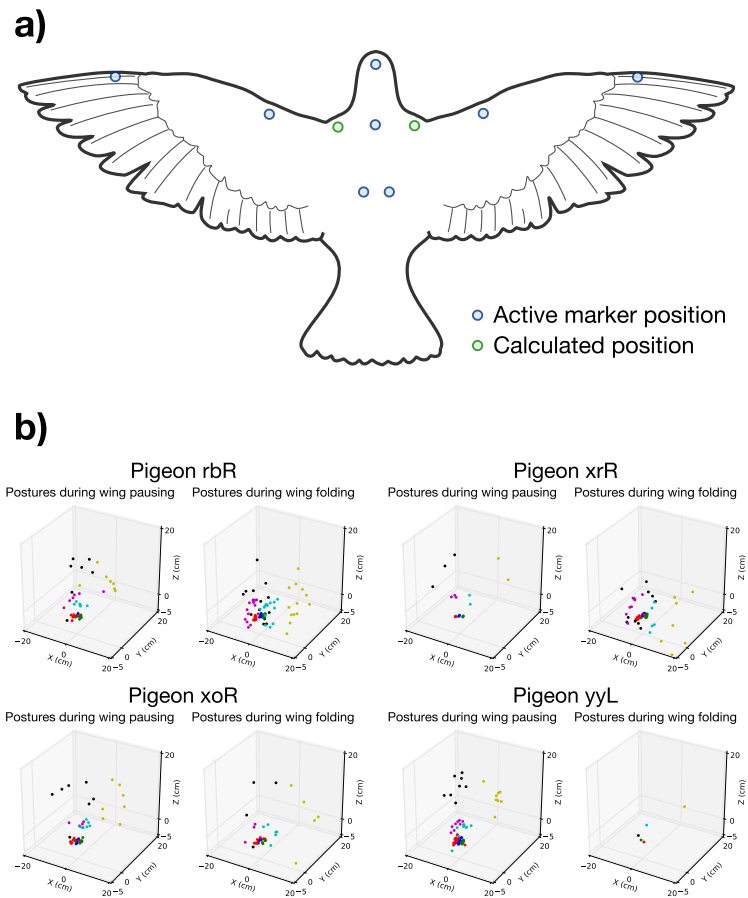

## Supplemental Figure 3

**Marker and anatomical locations** - Active infrared LED markers were attached to **a)** key anatomical locations to speed post-processing of flight videos. These markers appeared on camera as bright point sources. Markers were placed on the crown of the head, 2 cm from the left and right wingtips, at the left and right wrists, and in a plane on the bird's back/torso. For use in the rigid body models of the pigeon, some anatomical locations were back-calculated from marker positions and measured wing segment lengths. Elbow locations were not calculated or used, with the upper arm from shoulder to wrist being treated as a single unit for the static postures used to calculate response to perturbation. Shoulder locations, however, were calculated from measured shoulder widths, as in the plane of the back with the superior backpack marker directly between the shoulders. In **b)** the point clouds formed by these markers are shown for each individual, separated into the wings paused posture and the wings folded posture. Points are colored by location (upper back marker in red, lower left back marker in green, lower right back marker in blue, left wrist marker in cyan, right wrist marker in magenta, left wingtip marker in yellow, and right wingtip marker in black) and are shown relative to the mean of the back markers. The greater variability of the folded posture is shown in the more disperse clouds generated by its markers.